\documentclass[letterpaper]{article} 
\usepackage[submission]{aaai25}  
\usepackage{times}  
\usepackage{helvet} 
\usepackage{courier}  
\usepackage[hyphens]{url}  
\usepackage{graphicx} 
\usepackage{natbib}  
\usepackage{caption} 
\usepackage{algorithm}
\usepackage{algorithmic}

\usepackage{newfloat}
\usepackage{listings}
\DeclareCaptionStyle{ruled}{labelfont=normalfont,labelsep=colon,strut=off} 
\floatstyle{ruled}
\newfloat{listing}{tb}{lst}{}
\floatname{listing}{Listing}

\title{Verification methods for international AI agreements}
\author{
   Akash R. Wasil\textsuperscript{\rm 1,\rm 2}, Tom Reed\textsuperscript{\rm 2}, Jack William Miller\textsuperscript{\rm 2} and Peter Barnett\textsuperscript{\rm 3}
}
\affiliations{
    \textsuperscript{\rm 1}Georgetown University\\
    \textsuperscript{\rm 2} University of Cambridge, ERA AI Fellowship \\
    \textsuperscript{\rm 3} Independent
}

\usepackage{cleveref}
\crefname{figure}{Figure}{Figures}

\usepackage{xcolor}
\usepackage{tcolorbox}
\usepackage{enumitem}

\begin{document}

\onecolumn

\vspace*{\fill}

\begin{center}
    {\huge \textbf{Governing dual-use technologies: \\ Case studies of international security agreements \& \\ lessons for AI governance}}
    \vspace{1em}

    \textbf{Akash R. Wasil}$^{1,2}$, 
    \textbf{Peter Barnett}$^{3}$,
    \textbf{Michael Gerovitch}$^{2}$,\\
    \textbf{Roman Hauksson}$^{4}$,
    \textbf{Tom Reed}$^{2}$, 
    \textbf{Jack William Miller}$^{2}$
     \\

    \renewcommand{\thefootnote}{\arabic{footnote}}
    \setcounter{footnote}{0}
    
    $^{1}$Georgetown University (\texttt{aw1404@georgetown.edu})\\

    $^{2}$University of Cambridge, ERA AI Fellowship \\
    
    $^{3}$Independent \\

    $^{4}$University of Texas at Dallas
    
    \vspace{2em}
\end{center}

\begin{abstract}
\centering
\begin{minipage}[t]{0.8\textwidth}

\Large

International AI governance agreements and institutions may play an important role in reducing global security risks from advanced AI. To inform the design of such agreements and institutions, we conducted case studies of historical and contemporary international security agreements. We focused specifically on those arrangements around dual-use technologies, examining agreements in nuclear security, chemical weapons, biosecurity, and export controls. For each agreement, we examined four key areas: (a) purpose, (b) core powers, (c) governance structure, and (d) instances of non-compliance. From these case studies, we extracted lessons for the design of international AI agreements and governance institutions. We discuss the importance of robust verification methods, strategies for balancing power between nations, mechanisms for adapting to rapid technological change, approaches to managing trade-offs between transparency and security, incentives for participation, and effective enforcement mechanisms.
\end{minipage}
\end{abstract}

\vspace*{\fill}

\newpage

\section{\huge Executive Summary}
\vspace{1em}

{
\setlength{\parindent}{0pt}
\large
We examine 5 international security agreements focusing on dual-use technologies and extract lessons for potential AI governance efforts. The following table summarizes the 5 case studies:

\begin{tcolorbox}[title=Case Study Summaries]
\textbf{{\large IAEA (International Atomic Energy Agency)}}
\begin{itemize}
\item \textbf{Purpose}: Promote peaceful nuclear energy, prevent nuclear proliferation, and verify compliance with the Treaty on the Non-proliferation of Nuclear Weapons.
\item \textbf{Core Powers}: Inspect facilities, verify non-proliferation, and report violations to the UN Security Council.
\item \textbf{Governance}: The Board of Governors has 35 members --- 13 are pre-selected based on nuclear capability, and 22 are elected by the General Conference for two years. The Director General is chosen by the Board and approved by the General Conference.
\item \textbf{Non-Compliance}: Iran's illegal nuclear program was discovered in 2002, leading to sanctions and the 2015 Iran nuclear deal. The deal's effectiveness was questioned and undermined when the US withdrew.
\end{itemize}

\textbf{{\large START Treaties}}
\begin{itemize}
\item \textbf{Purpose}: Reduce US and Russian nuclear arsenals.
\item \textbf{Core Powers}: Allow on-site inspections, satellite monitoring, and data exchange.
\item \textbf{Governance}: Overseen by the Bilateral Consultative Commission (BCC) with US and Russian officials; decisions are made by consensus.
\item \textbf{Non-Compliance}: In 2023, Russia stopped allowing inspections.
\end{itemize}

\textbf{{\large The Organisation for the Prohibition of Chemical Weapons (OPCW)}}
\begin{itemize}
\item \textbf{Purpose}: Eliminate the production and use of chemical weapons and verify compliance with the Chemical Weapons Convention (CWC).
\item \textbf{Core Powers}: The OPCW conducts inspections and oversees the elimination of chemical destruction.
\item \textbf{Governance}: The Conference of the States Parties elects a 41-member Executive Council (EC) to the OPCW. The Director-General manages daily operations.
\item \textbf{Non-Compliance}: Syria used chemical weapons and failed to comply with OPCW inspections, leading to suspension from certain rights and privileges afforded by the CWC. 
\end{itemize}

\textbf{{\large Wassenaar Arrangement}}
\begin{itemize}
\item \textbf{Purpose}: Promote transparency in the exports of arms and dual-use technologies \& generate consensus.
\item \textbf{Core Powers}: Coordinate arms control policies and share best practices.
\item \textbf{Governance}: Decisions are made by consensus at annual meetings. Discussions are facilitated through working groups for technical matters and policy decisions.
\item \textbf{Non-Compliance}: Russia has been accused of undermining and violating the Wassenaar Arrangement by unilaterally blocking attempts to update export control lists.
\end{itemize}

\textbf{{\large Biological Weapons Convention (BWC)}}
\begin{itemize}
\item \textbf{Purpose}: Eliminate the production and use of biological weapons.
\item \textbf{Core Powers}: The BWC relies on commitments by member states and voluntary information-sharing mechanisms between nations (rather than a centralized body like the IAEA or the OPCW).
\item \textbf{Governance}: Decision-making is decentralized. States conduct Review Conferences to discuss new scientific developments. States address potential violations through consultation and cooperation; there is no formal process for inspections or verification.
\item \textbf{Non-Compliance}: The Soviet Union secretly developed biological weapons until exposed by defectors. The BWC failed to detect this violation, exposing potential weaknesses in the BWC relating to lack of verification and an over-reliance on self-reporting.
\end{itemize}
\end{tcolorbox}

\newpage

Drawing from these case studies, we extracted lessons learned that could inform international AI governance agreements and institutions:

\vspace{0.75em}

\begin{tcolorbox}[title=Key Findings and Lessons Learned]
\begin{enumerate}
\item \textbf{Verification mechanisms are essential:} Robust inspection and monitoring powers have been vital for institutions like the IAEA and OPCW, while the lack of verification in the BWC has undermined its effectiveness. Given AI's potential strategic value, effective verification will likely be a necessary foundation for any international agreement.

\item \textbf{Governance structures attempt to distribute power based on geography and national capabilities:} International agreements often provide extra power (e.g., some seats on the IAEA Board of Governors are reserved for nations with advanced nuclear programs) to nations with advanced capabilities while also attempting to ensure geographic representatives (e.g., some seats on the IAEA Board of Governors are reserved for nations from certain geographical regions). Future AI governance structures will need to consider how to appropriately represent global interests while acknowledging the roles of leading AI powers.

\item \textbf{Agreements balance transparency and privacy.} Greater transparency allows for more robust verification, though greater transparency could expose proprietary information or state secrets. The US rejected a proposed BWC verification method due to concerns about proprietary information. For AI agreements, the level of transparency nations are willing to allow will likely be proportional to the perceived risks. 

\item \textbf{Agreements must adapt to technological change:} The NEW START treaty and the BWC have had to adapt to novel technologies like hypersonic weapons or advances in synthetic biology. Rapid technological advancement in AI may require governance structures that can adapt quickly. This emphasizes the importance of strong technical expertise within governing bodies to track breakthroughs, anticipate risks, and develop appropriate regulatory responses.

\item \textbf{Agreements use benefit-sharing to incentivize participation.} The OPCW provides member states with support for the peaceful uses of chemistry and the IAEA provides support for the peaceful use of nuclear energy. International AI agreements may incentivize participation by sharing safe and responsible applications of AI technology.

\end{enumerate}
\end{tcolorbox}

\vspace{0.75em}

\subsection{\LARGE Future work}
These findings underscore the complexity of developing effective international AI governance mechanisms. Some promising areas for future work include:

\begin{itemize}
\item \textbf{Verification methods.} Developing robust, adaptable verification methods for AI development and deployment.
\item \textbf{Core powers and decisions.} Identifying key decisions that an international AI governance institution should make and how these decisions would be made.
\item \textbf{Governance structures.} Designing governance structures that balance global representation with the interests of leading AI nations. 
\item \textbf{Technical expertise.} Identifying ways to incorporate strong technical expertise into governance bodies to keep pace with rapid AI advancements.
\item \textbf{Non-compliance.} Examining strategies that could be used to prevent non-compliance or react to confirmed instances of non-compliance with AI agreements.
\end{itemize}

By drawing insights from case studies of international security agreements, we may be able to identify ways to make international AI agreements more effective and robust.
}

\twocolumn

\section{Introduction}

The development of advanced AI presents important global security risks. AI experts, policy experts, and world governments have acknowledged a variety of ways that advanced AI development could present major security challenges \citep{bengio-managing-extreme-risks, roberts2023governing, biden2023executive}. At the UK AI Safety Summit, over 20 nations (including the United States and China) acknowledged risks from the intentional misuse or unintentional misalignment of advanced AI systems. Furthermore, they acknowledged that ``[m]any risks arising from AI are inherently international in nature, and so are best addressed through international cooperation'' \citep{bletchley2023declaration}.

International coordination may have the potential to reduce global security risks. Thus far, international coordination has focused on defining risks and establishing voluntary standards for safe AI development. Many countries have launched AI Safety Institutes designed to measure risks from advanced AI, further the science of AI evaluations and risk management, share research about AI safety, and work toward a common understanding of risks and risk mitigation strategies \citep{ukgov2024aisafety}. The US AI Safety Institute announced that it plans to lead an ``inclusive, international network on the science of AI safety”~\citep{NIST2024usaisi}, and international dialogues have emphasized the need for “coordinated global action on AI safety research and governance'' \citep{idais2023, nist2024aisi}. Furthermore, while it is too early to tell how China will react to international coordination proposals, there are early signs that Chinese leaders acknowledge global security risks from advanced AI and could be interested in international approaches to AI governance \citep{wasil2024understanding}. 

Previous work has examined proposals for international AI agreements. Such agreements focus on various aims, including consensus-building, enforcement of regulations, emergency preparedness and response, and the shared distribution of benefits from AI \citep{maas2023international}. Some scholars have described an international approach to advanced AI development, in which certain kinds of advanced AI development take place in a joint safety-focused AI project \citep{cigiframework,hausenloy2023multinational, ho2023international}. Another common proposal involves the creation of international institutions that set standards, monitor compliance with standards, detect unauthorized AI development, or certify national regulatory bodies \citep{ho2023international, trager2023international}. 

International proposals for AI governance could be informed by existing international agreements and international institutions. There are many historical and modern case studies of international agreements or institutions that attempt to minimize global risks \citep{UN1945Charter, history-of-iaea, goldblat1997biological, croddy2002chemical, thakur2007chemical,  schabas2011introduction, woolf2011new, lipson2017wassenaar}. In this paper, we review international agreements which focus on dual-use technologies. These agreements are especially relevant to AI given its dual-use potential \citep{Pedersen_2023, NTIA2024}. Our aim is to identify lessons learned that could inform the design of future international agreements or international institutions designed to reduce global risks from advanced AI. The paper is divided into two sections:
\begin{enumerate}
    \item \textbf{Case studies.} We cover the following questions for each international agreement:
    \begin{enumerate}
        \item \textit{Purpose.} What is the purpose of the agreement?
        \item \textit{Core powers.} What are the core powers granted to the international body responsible for monitoring and enforcement?
        \item \textit{Governance structure.} How is the agreement governed? If an institution is established, how is decision-making power allocated between member nations?
        \item \textit{Case study of non-compliance.} Are there any cases in which a nation was suspected of non-compliance with the international agreement? How did this situation get resolved?
    \end{enumerate}
    \item \textbf{Lessons learned.} We summarize lessons learned that could be useful when thinking about international agreements related to the development or deployment of advanced AI.
\end{enumerate}

\section{Case studies of international agreements}

\subsection{IAEA}

\textbf{Purpose.} The International Atomic Energy Agency (IAEA) was founded in 1957 to promote the peaceful use of nuclear energy and limit its use for military purposes \citep{goldschmidt1977origins}. The IAEA exists as an autonomous organization within the United Nations. In practice, one of its main purposes is to verify that states do not build nuclear weapons \citep{scheinman2016international}. 

\textbf{Core powers.} The IAEA can conduct inspections to ensure that states are not secretly building nuclear facilities. Findings from inspections are reported to the IAEA Board of Directors. If the Board of Directors believes that a state is not complying with international agreements, the Board can escalate the issue to the UN Security Council \citep{rockwood2013legal}. 

Ultimately, the IAEA does not have the authority to take action directly --- it simply provides information to the UN Security Council and member states \citep{rockwood2013legal}. The UN Security Council has the authority to impose sanctions\footnote{After the UN Security Council passes a resolution to issue a sanction, each member state is responsible for implementing the sanctions at the national level. Member states must report back to the UN about their compliance and enforcement measures.} (e.g., trade restrictions, travel bans, freezing assets) or engage in military action. 

\begin{figure}[ht!]
    \centering
    \includegraphics[width=1\linewidth]{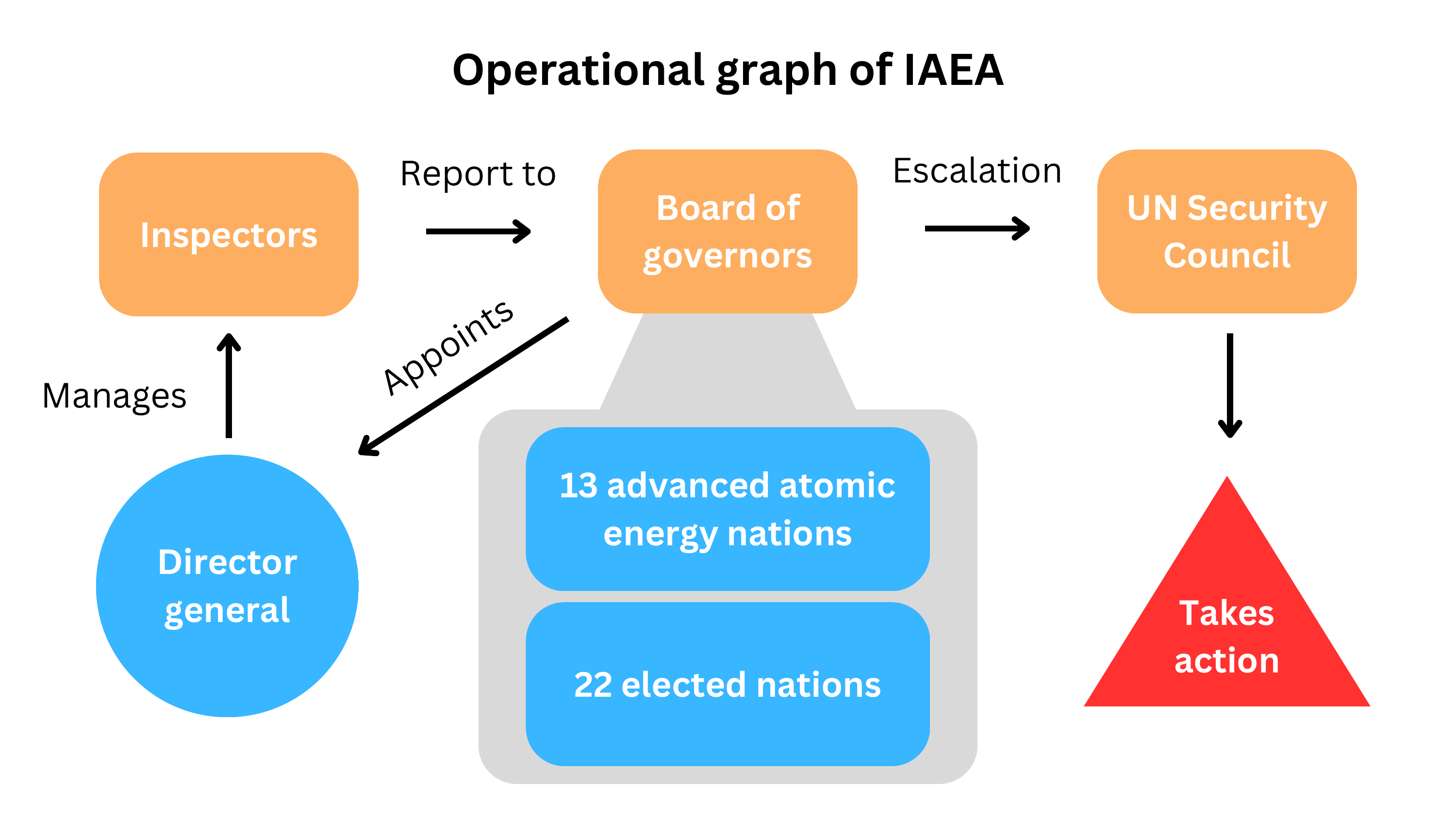}
    \caption{The governance structure of the IAEA.}
    \label{fig:iaea-governance-structure}
\end{figure}

\textbf{Governance structure.} The IAEA is composed of 178 member states \citep{IAEA_Member_States}. The selection of the Board of Governors and the Director General are the most consequential parts of IAEA’s governance structure. The Board currently consists of 35 members; each member represents a different nation \citep{iaea_board_of_governors}. 13 of those spots are guaranteed to nations that have advanced atomic energy technology 22 are elected by the General Conference for a two-year term. The election attempts to ensure a balanced geographical representation from each region of the world. The Director General of the IAEA is selected by the Board. The Director General must receive a two-thirds majority from the Board, as well as approval from the General Conference \citep{iaea_grossi_reappointment}. The Director General is responsible for setting the IAEA’s strategic direction, overseeing the development and implementation of policies, managing the IAEA’s staff, providing reports to the Board of Generals and General Conference, and directing the IAEA’s emergency response protocols. The Director General appoints many of the senior-level staff of the IAEA and oversees the recruitment and training of IAEA inspectors \citep{IAEA_DG, history-of-iaea}. 

\textbf{Non-compliance.} In the late 1990s and early 2000s, Iran began secretly investing in its nuclear program via the Amad Plan. In 2002, these efforts were publicly exposed by the National Council of Resistance of Iran (NCRI), a political organization that advocated for the collapse of the Islamic Republic and the establishment of a democratic, secular government \citep{NCI_revelations}. After the NCRI revelations, the IAEA conducted inspections and found Iran to be in violation of its obligations under the Nuclear Non-Proliferation Treaty \citep{gerami2012international}. 

The IAEA reported Iran to the UN Security Council who then recommended the application of diplomatic pressure and economic sanctions \citep{suzuki2019iran}. This led to a series of sanctions by the UN Security Council and a series of international negotiations with Iran. In 2015, this resulted in the Joint Comprehensive Plan of Action (JCPOA) --- an agreement between Iran and the P5+1 (US, UK, France, Russia, China, and Germany)~\citep{abtahi2014joint}. The JCPOA (colloquially referred to as the “Iran Nuclear Deal”) required Iran to limit its investments in its nuclear program and allow extensive monitoring and inspections by the IAEA. In exchange, economic sanctions on Iran would be lifted \citep{CFR_nucleardeal}.

The JCPOA went into effect in 2016, but the United States withdrew from the agreement in 2018 \citep{fitzpatrick2017assessing}. The Trump Administration argued that the JCPOA did not go far enough in restricting Iran’s nuclear program, did not restrict its ballistic missile program, and did not address its role in regional conflicts in the Middle East \citep{CFR_nucleardeal}. There were also concerns about the verification methods, with some critics of the deal arguing that Iran might still be able to secretly invest in its nuclear program despite the increased monitoring. Ultimately, the United States ended up reimposing sanctions on Iran. Iran initially continued to comply with the agreement (as verified by IAEA reports) \cite{davenport2019iaea}, but as the sanctions from the United States intensified, Iran began to reduce its compliance with the JCPOA commitments \cite{tabatabai2019iranian}. IAEA inspections continued to occur \cite{adler2010iran}, and the JCPOA did not fully dissolve– many nations (including Russia and China) have continued to support it, and there are ongoing efforts to bring the United States back into the deal.  

\subsection{START Treaties}

\textbf{Purpose.} The START treaties, or Strategic Arms Reduction Treaties, were a series of bilateral agreements between the United States and the Soviet Union (later Russia) \citep{schenck2011start}. The primary goal of the various START treaties is to reduce the number of strategic nuclear weapons and their delivery systems possessed by the United States and Russia (previously the Soviet Union). The treaties include various verification and transparency mechanisms to provide assurance to each nation that the other is meeting its commitments. START I entered into force in 1994. START I was intended to be followed by START II, but START II never entered into force due to disagreements between the two nations. SORT (Strategic Offensive Reductions Treaty) successfully entered into force in 2003, and was superseded by New Start in 2011 \citep{woolf2011new}.

\textbf{Core powers.} The original START I allowed nations to conduct inspections to ensure compliance, these included routine inspections and short-notice inspections of suspect activities \citep{woolf2011new}. The nations also monitored each other primarily using satellites and agreed not to interfere with each other’s monitoring. START I also allowed for continuous monitoring of certain facilities– inspectors from the other nation would maintain a 24/7 presence and inspect all vehicles and containers large enough to hold treaty-limited items. The current New START treaty allows for similar monitoring and verification by each nation. Unlike START I, New START does not include continuous on-site monitoring at missile production facilities, instead relying more on periodic inspections and data exchanges \citep{woolf2011new}. This may be due to continuous monitoring being expensive to maintain, and the greater level of trust between the U.S. and Russia. 

\begin{figure}[ht!]
    \centering
    \includegraphics[width=1\linewidth]{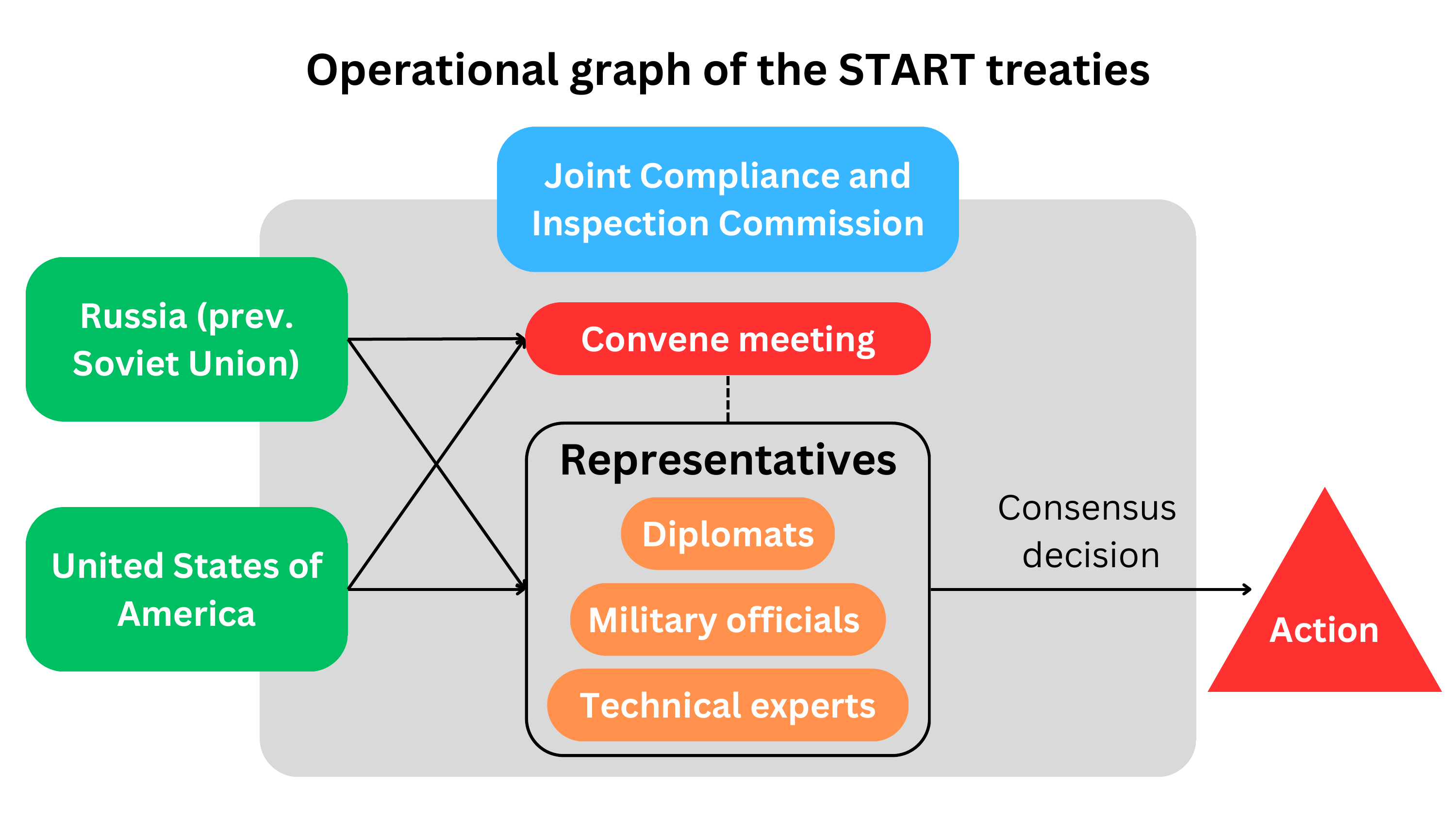}
    \caption{The governance structure of the START treaties.}
    \label{fig:start-governance-structure}
\end{figure}

\textbf{Governance structure.} START I created the Joint Compliance and Inspection Commission (JCIC), which was the primary forum for addressing implementation issues and resolving questions related to compliance \citep{schenck2011start}. The JCIC was composed of representatives from both the US and the USSR (later Russia). These representatives included diplomats, military officials, and technical experts. The JCIC met regularly, typically several times a year, and could convene special sessions at the request of either party. Decisions were made by consensus, and both parties had to agree on resolutions and interpretations. Proceedings were confidential, and agreed outcomes were shared with the relevant agencies of both countries. Both countries also authorized national implementation bodies to oversee the implementation of START I. The US body was the On-Site Inspection Agency (OSIA), which was originally created in 1988 to implement the inspection regime for the Intermediate-Range Nuclear Forces Treaty (INF). The OSIA trained inspectors, organized inspections of Russian facilities, and escorted Russian inspectors around US facilities. In 1998 it was consolidated into the newly formed Defense Threat Reduction Agency, which continues to carry out similar duties for New Start. New START replaced the JCIC with the Bilateral Consultative Commission (BCC), which has a similar function and is composed of representatives from the US and Russia \citep{woolf2011new}. Decisions are again made by consensus. 

\textbf{Non-compliance.} In January 2023, the U.S. State Department reported to Congress that Russia was in non-compliance with the New START treaty. The report noted that Russia had refused to reschedule inspections after their COVID-19-related pause and had failed to meet for a session of the Bilateral Consultative Commission since October 2021. This marked the first formal U.S. accusation of Russian violation since the treaty's inception in 2011. (Previously, there had been disputes with START I, although these had been successfully resolved via diplomatic channels and the JCIC.) 
Following this report, in February 2023, Russia announced the suspension of its participation in the New START treaty. This action escalated the existing compliance issues, as Russia had already halted on-site inspections in August 2022. Russian President Vladimir Putin cited Western support for Ukraine as the primary reason for the suspension. The move raised concerns about the future of data exchanges and other verification measures required by the treaty. In response, the US revoked visas to Russian inspectors in June 2023. 
Despite the suspension, both countries stated they would continue to adhere to the numerical limits on nuclear warheads and delivery systems set by the treaty. However, the lack of verification measures complicated the ability to confirm compliance. As of 2024, Russia has rejected further talks, continuing to cite US support of Ukraine. Without further negotiations, New START (along with its verification measures) will expire in 2026. 

\subsection{The Organisation for the Prohibition of Chemical Weapons (OPCW)}

\textbf{Purpose.} The Organisation for the Prohibition of Chemical Weapons (OPCW) is an autonomous international organization established to implement the Chemical Weapons Convention (CWC), a chemical weapons control treaty that went into force in 1997 \citep{OPCW_CWC}. The OPCW now oversees 193 member states that have signed the CWC. Under OPCW verification, members of the CWC are prohibited from developing, producing, stockpiling, transferring, or using chemical weapons (except for limited uses such as medical or research purposes) and are required to destroy any of their existing chemical weapons \citep{ACA_CWC}. The OPCW was modeled after the International Atomic Energy Agency (IAEA) \citep{dorn1993organization}.

\textbf{Core powers.} The OPCW has the authority to send inspectors to any member state to search for evidence of the production of banned chemicals and verify compliance with the CWC. OPCW conducts both routine inspections as well as investigations into allegations of CWC violations through Fact-Finding Missions (FFM) and challenge inspections. The organization also oversees and verifies the destruction of chemical weapons stockpiles and production facilities declared by member states.

If a Member State believes another state is non-compliant with the CWC, it can request the Executive Council (EC) to launch a challenge inspection. The inspection can be launched at short notice --- within 12 hours of notification --- and cannot be refused by the Member State. The Director-General formally issues the challenge inspection.

If a State Party fails to address compliance issues, the Conference may restrict or suspend its rights under the Convention. The Conference may also recommend collective measures, such as sanctions. For serious violations, the Conference may recommend collective measures to States Parties or bring the issue to the UN Security Council.

Another key power of the OPCW is to facilitate cooperation on safe chemistry research. While the CWC restricts the production of ‘dual-use’ chemicals, it encourages the peaceful uses of chemistry in industry, agriculture, and research purposes. The OPCW also offers training to specialists on practical aspects of chemical safety and provides forums to share and discuss best practices among State Parties \citep{cwc_opcw_1997}.

\begin{figure}[ht!]
    \centering
    \includegraphics[width=1\linewidth]{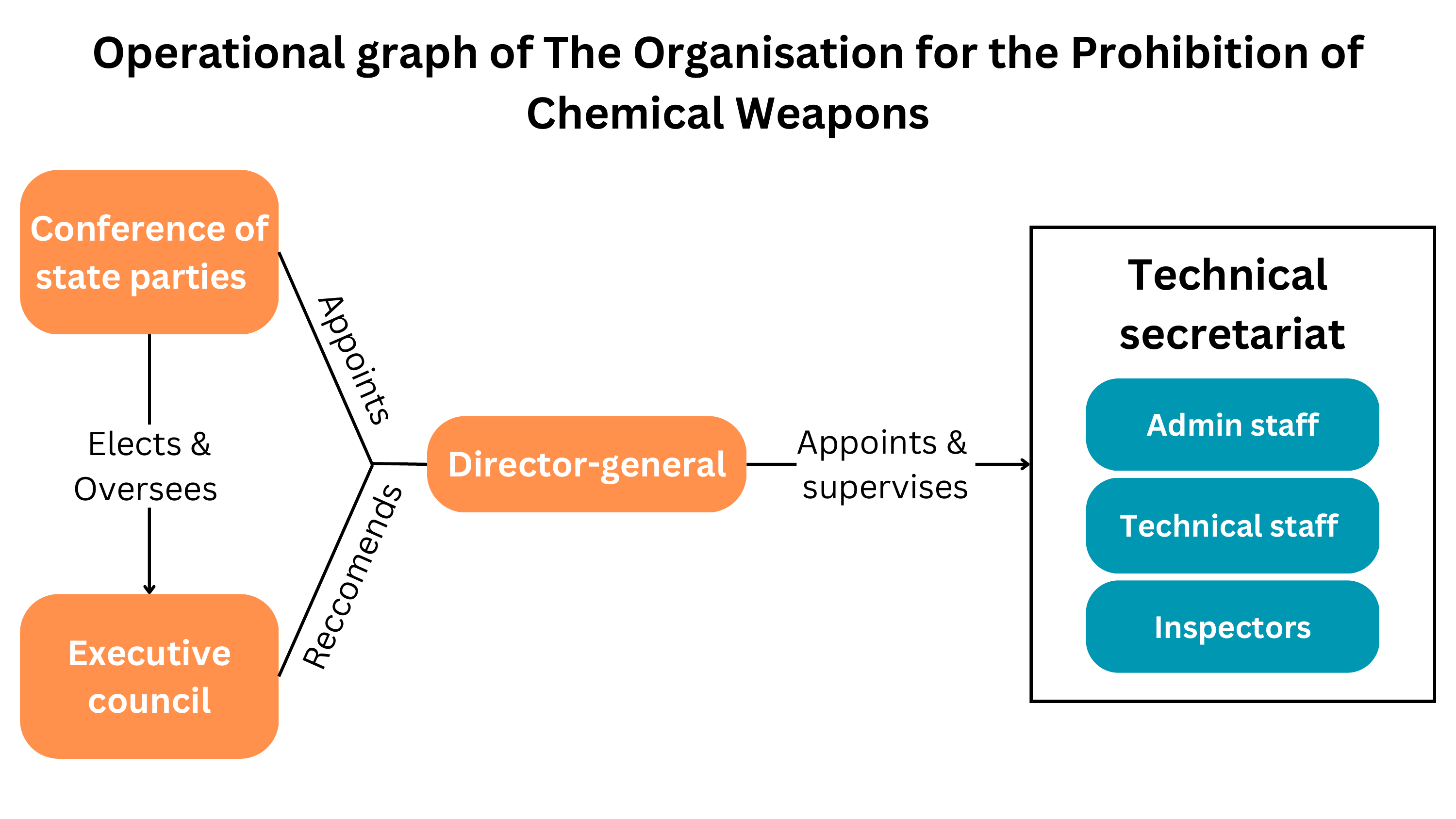}
    \caption{The governance structure of the chemical weapons convention. \textit{This figure is based on Figure 1 in \citet{dorn1995compliance}, originally sourced from \citet{paul1990disarmament}.}}
    \label{fig:chemical-weapons-structure}
\end{figure}

\textbf{Governance structure.} The Conference of the States Parties (CSP) is the principal organ composed of all OPCW Member States. It meets annually to make key decisions, adopt the budget, and elect and direct the Executive Council (EC), and jointly appoint the Director-General with the EC. Also, the States Parties nominate a group of emergency response experts to be part of the Protection Network, who are called in to assist and protect against chemical weapons.

The CWC mandates that each Member State establishes a National Authority to facilitate communication with the OPCW and ensure national compliance through appropriate legislation and enforcement measures. 

The Executive Council (EC), with 41 CSP-elected States Parties, oversees the day-to-day operation of OPCW, implements decisions of the CSP, and prepares recommendations for CSP. Under OPCW rules, the EC must have a fixed number of States Parties across geographic regions; a fixed subset of each region must be the States Parties with the most advanced national chemical industries in their region.

Within the EC, approving decisions generally requires a two-thirds majority. The exceptions are questions of procedure (which require a majority vote) and decisions to stop a challenge inspection within a 12-hour review period (which require a three-quarters majority vote.)

The Technical Secretariat, appointed and supervised by the Director-General, is responsible for the day-to-day operations of OPCW. If a State Party requests an inspection, the Director-General is responsible for notifying the CSP, including the challenged State Party, as well as the EC. The Director-General is also responsible for selecting the inspection team from a list of qualified experts. The Secretariat offers training for first responders, government experts, and emergency response units to support individual Member States in implementing the Convention \citep{cwc_opcw_1997}.

\textbf{Non-compliance.} Syria, despite ratifying the Chemical Weapons Convention (CWC) in 2013, was found to continue using chemical weapons and failed to comply with OPCW investigations. In April 2021, the OPCW suspended Syria's rights and privileges in the organization. Yet, Syria's ongoing non-compliance, Russia's diplomatic protection and vetoes at the UN Security Council, and OPCW's limited enforcement mechanisms have resulted in the situation in Syria remaining largely unresolved, with accountability for chemical weapons use still elusive  \citep{Masterson2020, UNSC_9372nd_Meeting_2023}. 

The United States determined that Russian forces had used chemical weapons against Ukrainian troops \citep{usstate_russia_disinformation_2024}. Russia lost its seat on the OPCW Executive Council during re-election and faced sanctions from the United States \citep{fdd_russia_opcw_2024}. While the OPCW commits to providing assistance and protection to Ukraine, there was no official OPCW inspection into Russia’s alleged violations since the evidence presented was ``insufficiently substantiated'' \citep{opcw_ukraine_statement_2024}.

\subsection{Wassenaar Arrangement}

\textbf{Purpose.} The Wassenaar Arrangement (WA) was established in 1996 as a voluntary export control regime \citep{davis1996wassenaar, wassenaar2024about}. Its primary purpose is to promote transparency in exports of conventional arms (such as tanks and missiles) and dual-use technologies (such as radio equipment and lasers) and to prevent accumulations of these items from destabilizing international security \citep{gartner2008wassenaar}. It is the successor to the Coordinating Committee for Multilateral Export Controls (CoCom), a stricter agreement that was established during the Cold War and ceased operation in 1994 \citep{Schiavone1997}.

\textbf{Core powers.} The Wassenaar Arrangement establishes regular information exchange and policy coordination, but it does not set up formal inspection or enforcement powers \citep{gartner2008wassenaar}. During the annual Plenary Meeting, all member states send representatives to a headquarters building in Vienna, and they come to a consensus on two “control lists” – the Munitions List and the List of Dual-Use Goods and Technologies – which define items that they agree should be subject to export controls \citep{wassenaar_guidelines_2016}. They also exchange information on transfers of these items and denials of certain export licenses.

The arrangement also maintains ``best practices'' documents on topics such as effective legislation on arms brokering and internal compliance programs
for dual-use goods and technologies \citep{wassenaar_guidelines_2016}. Each state retains its autonomy regarding whether or how it chooses to implement these practices --- as a result, each state's implementation of export controls differs in practice.

\begin{figure}[ht!]
    \centering
    \includegraphics[width=1\linewidth]{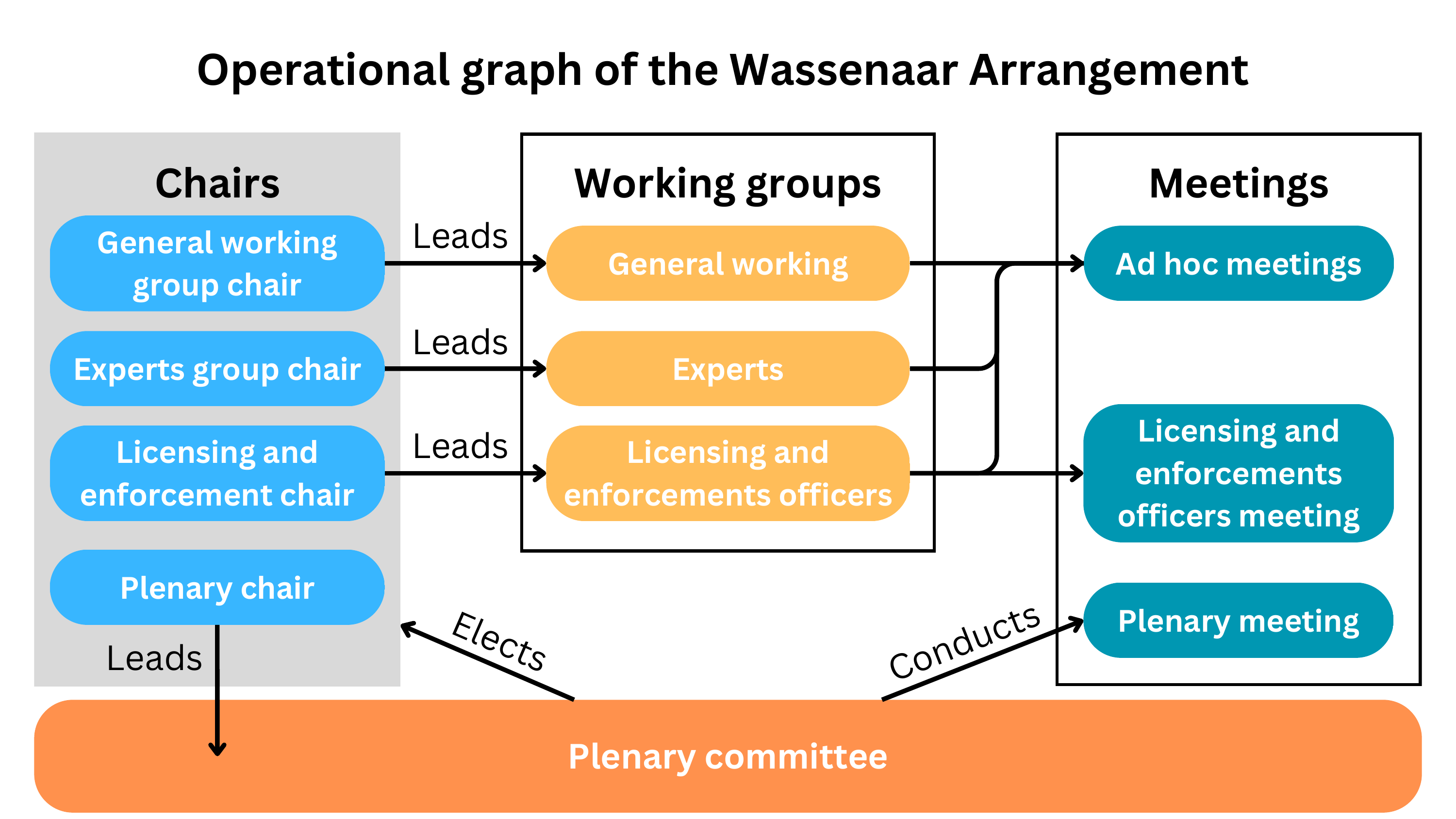}
    \caption{The governance structure of the Wassenaar Arrangement}
    \label{fig:wassenaar-arrangement}
\end{figure}

\textbf{Governance structure.} All decisions, including the addition of new members and the election of chairs, are made by consensus \citep{wassenaar2015basic}. The agreement is open to adding new members if they produce or export relevant goods, they maintain membership in other non-proliferation agreements, and they maintain fully effective export controls \citep{li2014cocom}.

The Plenary Committee is made up of representatives from all participating states and is headed by the Plenary Chair, who facilitates discussions at the annual Plenary Meeting in the headquarters building in Vienna. During the meeting, representatives provide information on transfers, revise best practices documents and control list entries, establish subsidiary ``working groups'' to help make recommendations for decisions, and decide on which state should be the Plenary Chair for the following year and who should lead each working group \citep{wassenaar2015basic}. The main working groups are currently the General Working Group (GWG), which deals with policy-related matters, and the Experts Group (EG), which addresses control lists. In addition, the Licensing and Enforcement Officers Meeting (LEOM) is held once per year, and a small group called the Secretariat provides administrative support and maintains the headquarters building.

\textbf{Non-compliance.} Since Russia's invasion of Ukraine, some experts on non-proliferation have questioned whether Russia can remain in export control arrangements such as the Wassenaar arrangement \citep{martin2022putins}. While their membership provides the international community some insight into their export activities, Russia has used its position to impede efforts to update control lists \citep{fact_sheet_wassenaar} forcing other nations to implement ad hoc export controls outside the arrangement's framework \citep{whitehouse2022fact}. Moreover, Russia exports parts used in weapons manufacturing to unstable regions, specifically North Korea \citep{herskovitz2024north}. Some view Russia's involvement in Wassenaar as an opportunity for intelligence gathering instead of a genuine attempt to further the goals of the arrangement \citep{fact_sheet_wassenaar}. Even if there were significant sentiment in favor of removing Russia from the Wassenaar arrangement, this wouldn't be feasible because some countries would oppose it --- all decisions are made by consensus --- and there is no formal expulsion mechanism.

\subsection{Biological Weapons Convention}

\textbf{Purpose.} The Biological Weapons Convention (BWC) prohibits the development, production, acquisition, transfer, stockpiling, and use of biological and toxin weapons \citep{kadlec1997biological}. It was the first multilateral disarmament treaty to ban an entire category of weapons of mass destruction. The BWC was opened for signature on April 10, 1972, and entered into force on March 26, 1975. It built upon the 1925 Geneva Protocol, which had only prohibited the use of biological weapons in war. The BWC currently has 187 state parties and four signatory states. The treaty consists of 15 articles. Review Conferences are held every five years to assess and strengthen the Convention's implementation.

\textbf{Core powers.} The BWC's core powers are primarily based on commitments by member states and information-sharing mechanisms, rather than a strong centralized authority. The Convention requires states to implement its provisions through national legislation and regulations. This includes prohibiting the development, production, and stockpiling of biological weapons, as well as destroying or diverting existing stockpiles to peaceful purposes. 

A key feature of the BWC is its system of Confidence Building Measures (CBMs), introduced in 1987. States parties are required to submit annual CBM reports by April 15th each year. These reports include information on research centers and laboratories, state biodefense programs, outbreaks of infectious diseases, relevant scientific publications, and vaccine production facilities. The CBMs aim to increase transparency and reduce suspicion among member states.
The BWC's implementation is supported by regular meetings. Review Conferences are held every five years to assess the Convention's effectiveness and consider new challenges. Since 2002, annual Meetings of States Parties and Meetings of Experts have been held to discuss specific topics related to the BWC's implementation.

Importantly, the Convention lacks a formal verification mechanism for compliance. In 1991, an ad-hoc group of experts (VEREX) was established to “identify and examine potential verification measures from a scientific and technical standpoint” \citep{bwc1991}. Ultimately, however, these efforts were unsuccessful, and the US rejected VEREX’s proposed protocol in 2001. 
Notably, the US’s rejection was grounded in the concern that the new proposals would not provide sufficient measures to effectively verify compliance with the protocol \citep{huigang2022development}. In the words of US Ambassador Mahley, the proposals would “still permit a potential proliferator to conceal significant efforts in legitimately undeclared facilities” \citep{whitehair2001bwc}. 

\begin{figure}[ht!]
    \centering
    \includegraphics[width=1\linewidth]{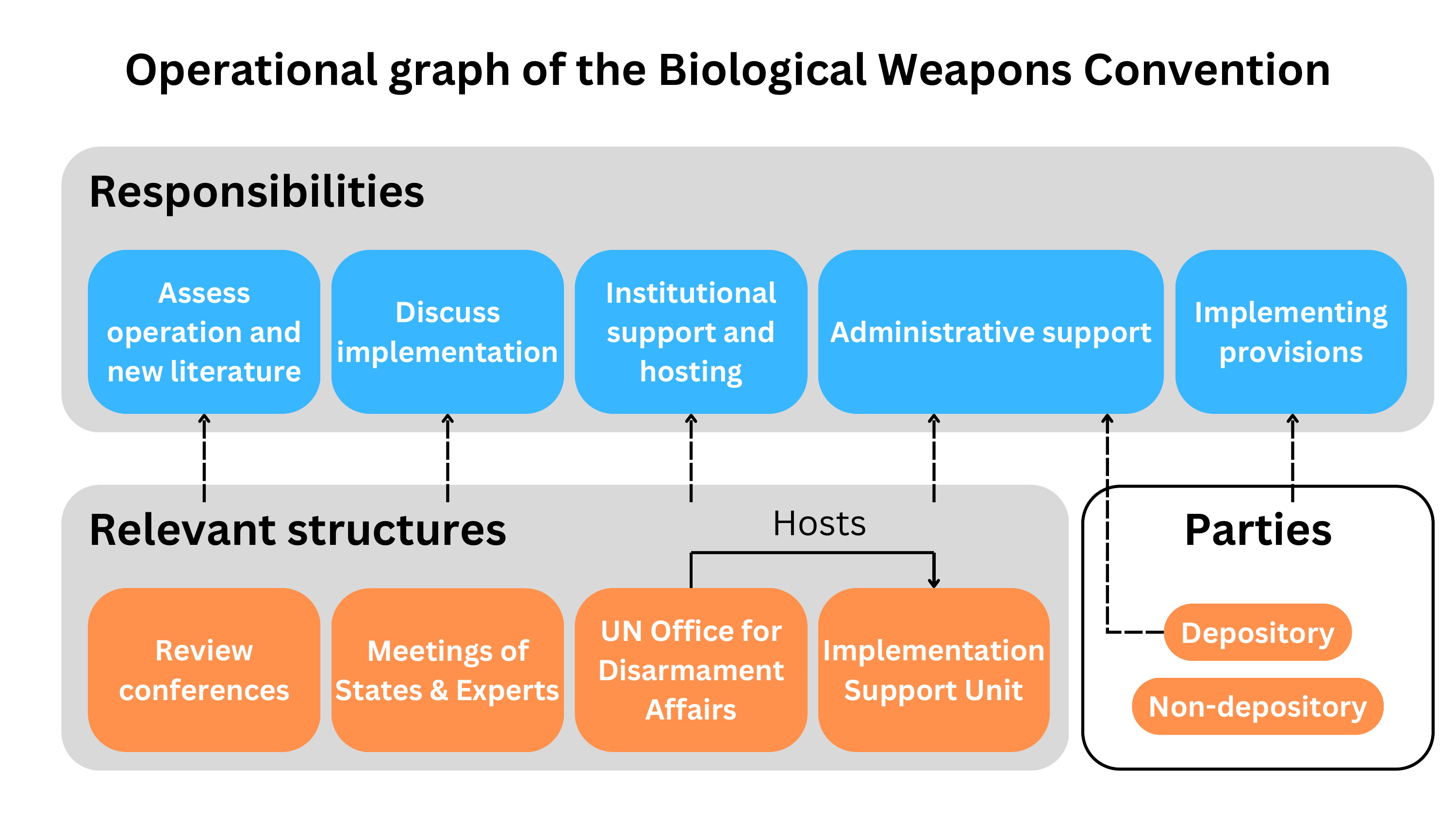}
    \caption{The governance structure of the Biological Weapons Convention.}
    \label{fig:bio-governance-structure}
\end{figure}

\textbf{Governance structure.} The BWC's governance structure is relatively decentralized, relying primarily on the collective action of its member states rather than a strong central authority. The key elements of its governance include:
\begin{enumerate}
    \item States Parties: All countries that have ratified the treaty. They are responsible for implementing the Convention's provisions through national legislation and regulations.
    \item Review Conferences: Held every five years, these conferences assess the operation of the Convention, consider new scientific and technological developments, and make decisions on further measures. The most recent was the Ninth Review Conference in November 2022.
    \item Meetings of States Parties and Meetings of Experts: Since 2002, these annual meetings have been held between Review Conferences to discuss specific topics related to the BWC's implementation.
    \item Implementation Support Unit (ISU): Established in 2006, this small unit of three full-time staff provides administrative support to States Parties, particularly in managing the Confidence Building Measures (CBMs) process.
    \item Depositary Governments: The US, UK, and Russia serve as depositary governments, responsible for certain administrative functions.
    \item United Nations Office for Disarmament Affairs (UNODA): Provides institutional support for the BWC, including hosting the ISU.
\end{enumerate}
The BWC does not have a formal international organization to oversee its implementation, unlike some other arms control treaties. Instead, it relies on the collective efforts of States Parties to monitor compliance and address concerns through consultation and cooperation.

\textbf{Non-compliance.} The Soviet Union’s Biopreparat program, operating from 1973 to 1991, is the most significant case of non-compliance with the BWC \citep{frischknecht2003history}. Despite being a signatory to the Convention, the Soviet Union established and maintained a large covert biological weapons program. Biopreparat employed over 50,000 people across various research and production facilities. They produced and stockpiled enormous quantities of deadly pathogens, including anthrax bacilli and smallpox virus. Some of these agents were even prepared for deployment via intercontinental ballistic missiles, demonstrating the program's integration with the Soviet strategic weapons complex \citep{roberts2003arms}.

The Biopreparat program remained secret for many years. Its existence only came to light after the defection of Vladimir Pasechnik to the UK in 1989, followed by Ken Alibek to the US in 1992 \citep{alibek1998behind, frischknecht2003history}. These defectors revealed the program's vast scope to Western intelligence agencies. This case highlights the potential for large-scale violations of the BWC to go undetected, especially in the absence of robust verification mechanisms. The BWC failed to detect this large-scale violation because of its lack of any robust verification measures and its reliance on self-reporting \citep{vertic2001}.

\section{Lessons learned}

In this paper, we reviewed international security agreements in an effort to understand how they are governed, what powers they possess, and how they handle issues of non-compliance. We reviewed agreements in nuclear security, chemical weapons security, biological weapons, and export controls. Some of these agreements involved the establishment of international institutions for monitoring and verification, while others relied on voluntary compliance between nations. Below, we discuss a few themes and lessons learned that could be useful for discussions about international AI governance. 

\textbf{Verification mechanisms are essential to assess compliance with international agreements.} The importance of robust verification mechanisms is highlighted by several case studies. The IAEA’s inspection powers for nuclear facilities, the OPCW’s authority to conduct challenge inspections for chemical weapons, and the on-site inspections or continuous monitoring in START treaties have played crucial roles in their effectiveness. In contrast, the lack of a formal verification protocol in the BWC permitted cases of non-compliance and undermined efforts to strengthen the treaty. 

International AI agreements will require robust verification methods to detect non-compliance. Strategies like on-site inspections, challenge inspections, and continuous monitoring could help ensure the robustness of verification regimes (see \citep{wasil2024verification}).

\textbf{Governance structures attempt to balance power between nations based on geography and geopolitical importance.} The international institutions we reviewed often had permanent or fixed seats for geopolitically powerful or technologically advanced nations, while also reserving a certain number of seats for countries from various regions around the world. For example, both the IAEA and the OPCW ensure representation in their membership whilst reserving additional or permanent powers to countries with advanced capabilities in the relevant fields. Bodies that rely strongly on consensus-based decision-making can lead to gridlock and limit an agreement’s adaptability, as in the Wassenaar Arrangement. Bilateral agreements (like the START treaties) can avoid gridlock but may break down in response to geopolitical events (such as Russia’s invasion of Ukraine). Furthermore, the willingness of key nations to follow through on commitments can be critical to the success of agreements– such as the United States withdrawing from the JCPOA. 

When considering AI agreements, it is important to consider the role of the United States and China– the world’s two leading AI powers. A bilateral agreement could attempt to draw from some of the promising provisions of the START treaties. A broader international agreement could attempt to ensure global representation while still preserving permanent seats or extra decision-making power for nations that lead in AI expertise. 

\textbf{Agreements require striking a balance between transparency and privacy.} Balancing the need for transparency with protecting legitimate state and commercial interests can be challenging \citep{chayes1998new}. This was evident in the U.S. rejection of the BWC verification protocol due to concerns about proprietary information \citep{vertic2001} and in the challenges faced by the Wassenaar Arrangement regarding agreeing on controlled items. 

For AI agreements, ensuring compliance with international standards or safety practices may require verification methods that promote high levels of transparency (e.g., on-site inspections, access to code, access to data centers) \citep{gallagher2003}. Depending on the perceived level of danger, nations may be more or less willing to tolerate invasive monitoring and verification methods. Furthermore, nations may justifiably want to secure sensitive or dangerous material, such as the model weights of highly dangerous systems or details about certain kinds of algorithmic insights. It will be important to identify verification methods that balance the need for transparency with other needs, such as security and national interests. Such verification methods could involve some techniques that are already standard for verifying compliance with international agreements, as well as novel approaches in which advanced hardware automatically alerts an international authority if it detects unauthorized code or unauthorized networking patterns \citep{kulp2024hardware}. 

\textbf{International institutions must adapt to rapid technological change.} The rapid pace of technological advancement poses challenges for international agreements. Both the NEW START treaty and the BWC have struggled to deal, respectively, with developing technologies like hypersonic weapons or advances in synthetic biology \citep{coetzee2021hypersonic}. International AI agreements will also have to adapt to technical breakthroughs. Examples include novel AI capabilities, new technical breakthroughs that make it easier for actors to develop dangerous AI, new technical breakthroughs that make it easier to monitor compliance with international agreements, and advances in AI that could allow AI to be incorporated into verification schemes. It will be essential for international AI governance institutions to possess strong technical expertise in order to track and incorporate such breakthroughs. The technical staff could play essential roles, such as interpreting model evaluations \citep{shevlane2023modelevaluationextremerisks}, evaluating affirmative safety cases \citep{clymer2024safety, wasil2024affirmative}, conducting interviews with technical experts to predict technical advances and anticipate security risks \citep{wasil2024understanding}, and identifying novel ways to detect non-compliance with international agreements (e.g., \citet{kulp2024hardware}). To acquire and retain such technical talent, international AI governance institutions may need to play more than merely a “policing” function– technical talent may be more attracted to projects that have a positive, inspiring, and innovative vision.\footnote{A related point was raised in the Acheson-Lilienthal Report, when the United States was considering establishing an international body to promote nuclear security. ``The difficulty of recruiting enforcement officers having only a negative and policing function, one of prohibiting, detecting, and suppressing, is obvious. Such a job lacks any dynamic qualities. It does not appeal to the imagination. Its future opportunities are obviously circumscribed. It might draw the kind of man, let us say, who was attracted to prohibition squads in years past. Compare this type of personnel with those who could be expected to enter a system under which it is clear that the constructive possibilities of atomic energy may also be developed. Atomic energy then becomes a new and creative field in which men may take pride as participants, whatever their particular role'' (Acheson-Lilienthal Report, 1946).}

\textbf{Agreements use benefit-sharing to incentivize participation.} Agreements often offer clear benefits to participating states in order to give them an incentive to join or remain \citep{barrett2007why}. Examples include the OPCW's support for peaceful uses of chemistry and the IAEA's promotion of peaceful nuclear energy. Future AI governance structures could consider incorporating mechanisms that promote beneficial AI research and development while mitigating risks.

\textbf{Enforcement is challenging and may rely on other national or international institutions.} Many international bodies lack direct enforcement powers. The IAEA and OPCW can only report violations to the UN Security Council, which then decides on enforcement actions. Reliance on external bodies for enforcement can lead to political deadlocks, as seen in the case of Syria's chemical weapons use. It will be important to consider what powers ought to be granted to a potential international AI governance institution. Example questions include: (a) what actions should it be allowed to take on its own, (b) to what extent will it rely on the UN Security Council to take actions, and (c) to what extent will it rely on individual member nations.

\bibliography{aaai25}

\begin{thebibliography}{80}
\providecommand{\natexlab}[1]{#1}

\bibitem[{Abtahi(2014)}]{abtahi2014joint}
Abtahi, H. 2014.
\newblock Joint Plan of Action on Iran’s Nuclear Program.
\newblock \emph{International Legal Materials}, 53(4): 732--738.

\bibitem[{Adler(2024)}]{adler2010iran}
Adler, M. 2024.
\newblock Iran and the IAEA.
\newblock \url{https://iranprimer.usip.org/resource/iran-and-iaea} Accessed: 2024-08-15.

\bibitem[{Alibek(1998)}]{alibek1998behind}
Alibek, D.~K. 1998.
\newblock Behind the mask: biological warfare.

\bibitem[{{Arms Control Association}(2024)}]{ACA_CWC}
{Arms Control Association}. 2024.
\newblock The Chemical Weapons Convention (CWC) at a Glance.
\newblock \url{https://www.armscontrol.org/factsheets/chemical-weapons-convention-cwc-glance-0}. Accessed 2024-08-14.

\bibitem[{Association(2020)}]{Masterson2020}
Association, A.~C. 2020.
\newblock Russia Disputes OPCW Findings.
\newblock \url{https://www.armscontrol.org/act/2020-03/news/russia-disputes-opcw-findings} Accessed 2024-08-2024.

\bibitem[{Barrett(2007)}]{barrett2007why}
Barrett, S. 2007.
\newblock \emph{Why Cooperate? The Incentive to Supply Global Public Goods}.
\newblock Oxford University Press.

\bibitem[{Bengio et~al.(2024)Bengio, Hinton, Yao, Song, Abbeel, Darrell, Harari, Zhang, Xue, Shalev-Shwartz et~al.}]{bengio-managing-extreme-risks}
Bengio, Y.; Hinton, G.; Yao, A.; Song, D.; Abbeel, P.; Darrell, T.; Harari, Y.~N.; Zhang, Y.-Q.; Xue, L.; Shalev-Shwartz, S.; et~al. 2024.
\newblock Managing extreme AI risks amid rapid progress.
\newblock \emph{Science}, 384(6698): 842--845.

\bibitem[{Biden(2023)}]{biden2023executive}
Biden, J.~R. 2023.
\newblock Executive order on the safe, secure, and trustworthy development and use of artificial intelligence.

\bibitem[{{Bletchley Declaration}(2023)}]{bletchley2023declaration}
{Bletchley Declaration}. 2023.
\newblock The Bletchley Declaration by Countries Attending the AI Safety Summit, 1-2 November 2023.

\bibitem[{{BWC}(1991)}]{bwc1991}
{BWC}. 1991.
\newblock Third Review Conference of the Parties to the Convention on the Prohibition of then Development, Production and Stockpiling of Bacteriological (Biological) and Toxin Weapons and on their Destruction.
\newblock BWC/CONT.111/23.

\bibitem[{Cass-Beggs et~al.(2024)Cass-Beggs, Clare, Dimowo, and Kara}]{cigiframework}
Cass-Beggs, D.; Clare, S.; Dimowo, D.; and Kara, Z. 2024.
\newblock Framework Convention on Global AI Challenges.

\bibitem[{{Center for Arms Control and Non-Proliferation}(2023)}]{fact_sheet_wassenaar}
{Center for Arms Control and Non-Proliferation}. 2023.
\newblock Fact Sheet: The Wassenaar Arrangement.

\bibitem[{Chayes and Chayes(1998)}]{chayes1998new}
Chayes, A.; and Chayes, A.~H. 1998.
\newblock \emph{The New Sovereignty: Compliance with International Regulatory Agreements}.
\newblock Harvard University Press.

\bibitem[{Clymer et~al.(2024)Clymer, Gabrieli, Krueger, and Larsen}]{clymer2024safety}
Clymer, J.; Gabrieli, N.; Krueger, D.; and Larsen, T. 2024.
\newblock Safety cases: Justifying the safety of advanced AI systems.
\newblock \emph{arXiv preprint arXiv:2403.10462}.

\bibitem[{Coetzee(2021)}]{coetzee2021hypersonic}
Coetzee, E. 2021.
\newblock Hypersonic weapons and the future of nuclear deterrence.
\newblock \emph{Scientia Militaria: South African Journal of Military Studies}, 49(1).

\bibitem[{Croddy, Perez-Armendariz, and Hart(2002)}]{croddy2002chemical}
Croddy, E.; Perez-Armendariz, C.; and Hart, J. 2002.
\newblock \emph{Chemical and biological warfare: a comprehensive survey for the concerned citizen}, volume~22.
\newblock Springer.

\bibitem[{Davenport(2019)}]{davenport2019iaea}
Davenport, K. 2019.
\newblock IAEA Says Iran Abiding by Nuclear Deal.
\newblock \url{https://www.armscontrol.org/act/2019-04/news/iaea-says-iran-abiding-nuclear-deal} Accessed: 2024-08-15.

\bibitem[{Davis(1996)}]{davis1996wassenaar}
Davis, L.~E. 1996.
\newblock The Wassenaar Arrangement.
\newblock \emph{Department of State Dispatch}, 7: 19.

\bibitem[{Dorn and Rolya(1993)}]{dorn1993organization}
Dorn, A.~W.; and Rolya, A. 1993.
\newblock The Organization for the Prohibition of Chemical Weapons and the IAEA: A comparative overview.
\newblock \emph{IAEA BULLETIN}, 35: 44--44.

\bibitem[{Dorn and Scott(1995)}]{dorn1995compliance}
Dorn, A.~W.; and Scott, D. 1995.
\newblock \emph{The Compliance Provisions in the Chemical Weapons Convention: A Summary and Analysis}, volume~95.
\newblock Graduate Institute of International Studies.

\bibitem[{Findlay and Meier(2001)}]{vertic2001}
Findlay, T.; and Meier, O., eds. 2001.
\newblock \emph{Verification Yearbook 2001}.
\newblock London: The Verification Research, Training and Information Centre (VERTIC).

\bibitem[{Fischer(1997)}]{history-of-iaea}
Fischer, D. 1997.
\newblock \emph{History of the International Atomic Energy Agency The first forty years}.
\newblock International Atomic Energy Agency (IAEA): IAEA.

\bibitem[{Fitzpatrick(2017)}]{fitzpatrick2017assessing}
Fitzpatrick, M. 2017.
\newblock Assessing the JCPOA.
\newblock \emph{Adelphi Series}, 57(466-467): 19--60.

\bibitem[{{Foundation for Defense of Democracies}(2024)}]{fdd_russia_opcw_2024}
{Foundation for Defense of Democracies}. 2024.
\newblock For Russia, a Year of Setbacks at the OPCW.
\newblock \emph{Foundation for Defense of Democracies Analysis}.
\newblock \url{https://www.opcw.org/media-centre/news/2024/05/statement-ukraine-opcw-spokesperson} Accessed: 2024-08-15.

\bibitem[{Frischknecht(2003)}]{frischknecht2003history}
Frischknecht, F. 2003.
\newblock The history of biological warfare: Human experimentation, modern nightmares and lone madmen in the twentieth century.
\newblock \emph{EMBO reports}, 4(S1): S47--S52.

\bibitem[{Gallagher(2003)}]{gallagher2003}
Gallagher, N.~W. 2003.
\newblock \emph{The Politics of Verification}.
\newblock Johns Hopkins University Press.

\bibitem[{G{\"a}rtner(2008)}]{gartner2008wassenaar}
G{\"a}rtner, H. 2008.
\newblock The Wassenaar Arrangement (WA): How it is broken and needs to be fixed.
\newblock \emph{Defence \& Security Analysis}, 24(1): 53--60.

\bibitem[{Gerami and Goldschmidt(2012)}]{gerami2012international}
Gerami, N.; and Goldschmidt, P. 2012.
\newblock \emph{The International Atomic Energy Agency's Decision to Find Iran in Non-Compliance, 2002-2006}.
\newblock National Defense University Press Washington, DC.

\bibitem[{Goldblat(1997)}]{goldblat1997biological}
Goldblat, J. 1997.
\newblock The biological weapons convention: An overview.
\newblock \emph{International Review of the Red Cross (1961-1997)}, 37(318): 251--265.

\bibitem[{Goldschmidt(1977)}]{goldschmidt1977origins}
Goldschmidt, B. 1977.
\newblock The Origins of the International Atomic Energy Agency.
\newblock \emph{IAEA Bulletin}, 19(4): 12.

\bibitem[{Hausenloy, Miotti, and Dennis(2023)}]{hausenloy2023multinational}
Hausenloy, J.; Miotti, A.; and Dennis, C. 2023.
\newblock Multinational AGI Consortium (MAGIC): A Proposal for International Coordination on AI.
\newblock \emph{arXiv preprint arXiv:2310.09217}.

\bibitem[{Herskovitz(2024)}]{herskovitz2024north}
Herskovitz, J. 2024.
\newblock North Korea and Russia Accelerate Exchange of Weapons and Resources.
\newblock \emph{Time}.
\newblock \url{https://time.com/6835478/north-korea-speed-up-weapons-shipment-russia/}. Accessed 2024-08-15.

\bibitem[{Ho et~al.(2023)Ho, Barnhart, Trager, Bengio, Brundage, Carnegie, Chowdhury, Dafoe, Hadfield, Levi et~al.}]{ho2023international}
Ho, L.; Barnhart, J.; Trager, R.; Bengio, Y.; Brundage, M.; Carnegie, A.; Chowdhury, R.; Dafoe, A.; Hadfield, G.; Levi, M.; et~al. 2023.
\newblock International institutions for advanced AI.
\newblock \emph{arXiv preprint arXiv:2307.04699}.

\bibitem[{Huigang et~al.(2022)Huigang, Menghui, Xiaoli, Cui, and Zhiming}]{huigang2022development}
Huigang, L.; Menghui, L.; Xiaoli, Z.; Cui, H.; and Zhiming, Y. 2022.
\newblock Development of and prospects for the biological weapons convention.
\newblock \emph{Journal of Biosafety and Biosecurity}, 4(1): 50--53.

\bibitem[{{IAEA}(2024)}]{IAEA_DG}
{IAEA}. 2024.
\newblock Director General's Office.
\newblock \url{https://www.iaea.org/about/organizational-structure/offices-reporting-to-the-director-general/director-generals-office}.

\bibitem[{{IDAIS}(2023)}]{idais2023}
{IDAIS}. 2023.
\newblock International Dialogues on {AI} Safety.

\bibitem[{{International Atomic Energy Agency}(2023{\natexlab{a}})}]{iaea_board_of_governors}
{International Atomic Energy Agency}. 2023{\natexlab{a}}.
\newblock Board of Governors.
\newblock \url{https://www.iaea.org/about/governance/board-of-governors} Accessed: 2024-08-15.

\bibitem[{{International Atomic Energy Agency}(2023{\natexlab{b}})}]{iaea_grossi_reappointment}
{International Atomic Energy Agency}. 2023{\natexlab{b}}.
\newblock IAEA Board of Governors Reappoints Director General Grossi for New 4-Year Term.
\newblock \url{https://www.iaea.org/newscenter/pressreleases/iaea-board-of-governors-reappoints-director-general-grossi-for-new-4-year-term} Accessed: 2024-08-15.

\bibitem[{{International Atomic Energy Agency}(2024)}]{IAEA_Member_States}
{International Atomic Energy Agency}. 2024.
\newblock List of Member States.
\newblock \url{https://www.iaea.org/about/governance/list-of-member-states}.

\bibitem[{{James Martin Center for Nonproliferation Studies}(2022)}]{martin2022putins}
{James Martin Center for Nonproliferation Studies}. 2022.
\newblock Putin’s War with Ukraine: Voices of CNS Experts on the Russian Invasion.

\bibitem[{Kadlec, Zelicoff, and Vrtis(1997)}]{kadlec1997biological}
Kadlec, R.~P.; Zelicoff, A.~P.; and Vrtis, A.~M. 1997.
\newblock Biological weapons control: prospects and implications for the future.
\newblock \emph{Jama}, 278(5): 351--356.

\bibitem[{Kulp et~al.(2024)Kulp, Gonzales, Smith, Heim, Puri, Vermeer, and Winkelman}]{kulp2024hardware}
Kulp, G.; Gonzales, D.; Smith, E.; Heim, L.; Puri, P.; Vermeer, M.~J.; and Winkelman, Z. 2024.
\newblock Hardware-Enabled Governance Mechanisms.

\bibitem[{Li(2014)}]{li2014cocom}
Li, M. 2014.
\newblock From CoCom to Wassenaar Arrangement and UNSCR 1540: A Historical Review of Multilateral High-tech Export Control Policy Development.
\newblock \emph{Available at SSRN 3664946}.

\bibitem[{Lipson(2017)}]{lipson2017wassenaar}
Lipson, M. 2017.
\newblock The Wassenaar Arrangement: Transparency and Restraint through Trans-Governmental Cooperation?
\newblock In \emph{Non-Proliferation Export Controls}, 57--82. Routledge.

\bibitem[{Maas and Villalobos(2023)}]{maas2023international}
Maas, M.~M.; and Villalobos, J.~J. 2023.
\newblock International AI institutions: A literature review of models, examples, and proposals.

\bibitem[{{National Telecommunications and Information Administration}(2024)}]{NTIA2024}
{National Telecommunications and Information Administration}. 2024.
\newblock Dual-Use Foundation Models with Widely Available Model Weights.
\newblock Technical report, U.S. Department of Commerce.

\bibitem[{NIST(2024)}]{NIST2024usaisi}
NIST. 2024.
\newblock The United States Artificial Intelligence Safety Institute: Vision, Mission, and Strategic Goals.
\newblock Technical report, U.S. Department of Commerce, Washington, D.C.

\bibitem[{{NIST}(2024)}]{nist2024aisi}
{NIST}. 2024.
\newblock The {United States} {Artificial Intelligence Safety Institute}: Vision, Mission, and Strategic Goals.
\newblock Technical report, National Institute of Standards and Technology.

\bibitem[{{Nuclear Control Institute}(2006)}]{NCI_revelations}
{Nuclear Control Institute}. 2006.
\newblock List of Revelations on Iran's Nuclear \& WMD Activities by the Iranian Opposition since 2002.
\newblock \url{https://www.nci.org/06nci/01-31/Revelations.htm} Accessed 2024-08-15.

\bibitem[{{OPCW}(2024)}]{OPCW_CWC}
{OPCW}. 2024.
\newblock Chemical Weapons Convention.
\newblock \url{https://www.opcw.org/chemical-weapons-convention}.

\bibitem[{{Organisation for the Prohibition of Chemical Weapons}(1997)}]{cwc_opcw_1997}
{Organisation for the Prohibition of Chemical Weapons}. 1997.
\newblock Convention on the Prohibition of the Development, Production, Stockpiling and Use of Chemical Weapons and on Their Destruction.
\newblock \url{{https://www.opcw.org/sites/default/files/documents/CWC/CWC_en.pdf}}.

\bibitem[{{Organisation for the Prohibition of Chemical Weapons}(2024)}]{opcw_ukraine_statement_2024}
{Organisation for the Prohibition of Chemical Weapons}. 2024.
\newblock Statement on Ukraine by OPCW Spokesperson.
\newblock \url{https://www.opcw.org/media-centre/news/2024/05/statement-ukraine-opcw-spokesperson} Accessed: 2024-08-15.

\bibitem[{Paul et~al.(1990)}]{paul1990disarmament}
Paul, D.; et~al. 1990.
\newblock \emph{Disarmament's Missing Dimension: A UN Agency to Administer Multilateral Treaties}, volume~1.
\newblock Dundurn.

\bibitem[{Pedersen(2023)}]{Pedersen_2023}
Pedersen, G.~O. 2023.
\newblock Note to Correspondents: Joint call by the United Nations Secretary-General and the President of the International Committee of the Red Cross for States to establish new prohibitions and restrictions on Autonomous Weapon Systems.
\newblock \url{https://www.un.org/sg/en/content/sg/note-correspondents/2023-10-05/note-correspondents-joint-call-the-united-nations-secretary-general-and-the-president-of-the-international-committee-of-the-red-cross-for-states-establish-new} Accessed: 2024-08-15.

\bibitem[{Roberts(2003)}]{roberts2003arms}
Roberts, G.~B. 2003.
\newblock \emph{Arms control without arms control: The failure of the biological weapons convention protocol and a new paradigm for fighting the threat of biological weapons}, volume~49.
\newblock USAF Institute for National Security Studies.

\bibitem[{Roberts et~al.(2023)Roberts, Cowls, Hine, Morley, Wang, Taddeo, and Floridi}]{roberts2023governing}
Roberts, H.; Cowls, J.; Hine, E.; Morley, J.; Wang, V.; Taddeo, M.; and Floridi, L. 2023.
\newblock Governing artificial intelligence in China and the European Union: Comparing aims and promoting ethical outcomes.
\newblock \emph{The Information Society}, 39(2): 79--97.

\bibitem[{{Robinson}(2023)}]{CFR_nucleardeal}
{Robinson}. 2023.
\newblock What Is the Iran Nuclear Deal?
\newblock \url{https://www.cfr.org/backgrounder/what-iran-nuclear-deal} Accessed 2024-08-15.

\bibitem[{Rockwood(2013)}]{rockwood2013legal}
Rockwood, L. 2013.
\newblock Legal framework for IAEA safeguards.

\bibitem[{Schabas(2011)}]{schabas2011introduction}
Schabas, W.~A. 2011.
\newblock \emph{An introduction to the international criminal court}.
\newblock Cambridge University Press.

\bibitem[{Scheinman(2016)}]{scheinman2016international}
Scheinman, L. 2016.
\newblock \emph{The international atomic energy agency and world nuclear order}.
\newblock Routledge.

\bibitem[{Schenck and Youmans(2011)}]{schenck2011start}
Schenck, L.~M.; and Youmans, R.~A. 2011.
\newblock From Start to Finish: A Historical Review of Nuclear Arms Controls Treaties and Starting over with the New Start.
\newblock \emph{Cardozo J. Int'l \& Comp. L.}, 20: 399.

\bibitem[{Schiavone(1997)}]{Schiavone1997}
Schiavone, G. 1997.
\newblock \emph{W. In: International Organizations.}, 289--306.
\newblock London: Palgrave Macmillan UK.

\bibitem[{Shevlane et~al.(2023)Shevlane, Farquhar, Garfinkel, Phuong, Whittlestone, Leung, Kokotajlo, Marchal, Anderljung, Kolt et~al.}]{shevlane2023modelevaluationextremerisks}
Shevlane, T.; Farquhar, S.; Garfinkel, B.; Phuong, M.; Whittlestone, J.; Leung, J.; Kokotajlo, D.; Marchal, N.; Anderljung, M.; Kolt, N.; et~al. 2023.
\newblock Model evaluation for extreme risks.

\bibitem[{Suzuki(2019)}]{suzuki2019iran}
Suzuki, K. 2019.
\newblock Iran: The role and effectiveness of UN sanctions.
\newblock In \emph{Economic sanctions in international law and practice}, 178--199. Routledge.

\bibitem[{Tabatabai and Pease(2019)}]{tabatabai2019iranian}
Tabatabai, A.; and Pease, C. 2019.
\newblock The Iranian Nuclear Negotiations.
\newblock \emph{How Negotiations End}, 27--45.

\bibitem[{Thakur and Haru(2007)}]{thakur2007chemical}
Thakur, R.; and Haru, E. 2007.
\newblock \emph{The Chemical Weapons Convention: Implementation, Challenges, Opportunities}.
\newblock Pearson Education India.

\bibitem[{{The White House}(2022)}]{whitehouse2022fact}
{The White House}. 2022.
\newblock Fact Sheet: Joined by Allies and Partners, the United States Imposes Devastating Costs on Russia.
\newblock Statements and releases.

\bibitem[{Trager et~al.(2023)Trager, Harack, Reuel, Carnegie, Heim, Ho, Kreps, Lall, Larter, h{\'E}igeartaigh et~al.}]{trager2023international}
Trager, R.; Harack, B.; Reuel, A.; Carnegie, A.; Heim, L.; Ho, L.; Kreps, S.; Lall, R.; Larter, O.; h{\'E}igeartaigh, S.~{\'O}.; et~al. 2023.
\newblock International governance of civilian AI: A jurisdictional certification approach.
\newblock \emph{arXiv preprint arXiv:2308.15514}.

\bibitem[{{UK Government}(2024)}]{ukgov2024aisafety}
{UK Government}. 2024.
\newblock Global leaders agree to launch first international network of {AI} {Safety} {Institutes} to boost cooperation of {AI}.

\bibitem[{{United Nations}(1945)}]{UN1945Charter}
{United Nations}. 1945.
\newblock Charter of the United Nations.
\newblock \url{https://www.un.org/en/about-us/un-charter}.
\newblock Signed in San Francisco, came into force on October 24, 1945. Amended in 1963, 1965, and 1973.

\bibitem[{{United Nations Security Council}(2023)}]{UNSC_9372nd_Meeting_2023}
{United Nations Security Council}. 2023.
\newblock {Syria’s Chemical Weapons Declaration Still Inaccurate, Unfinished, Top Disarmament Official Tells Security Council, Reiterating Need for Damascus to Fully Cooperate}.
\newblock \url{https://press.un.org/en/2023/sc15350.doc.htm}.
\newblock 9372nd Meeting (AM), SC/15350, 11 July 2023.

\bibitem[{{U.S. Department of State}(2024)}]{usstate_russia_disinformation_2024}
{U.S. Department of State}. 2024.
\newblock Russia Spreads Disinformation to Cover Up Its Use of Chemical Weapons in Ukraine.
\newblock \url{https://www.state.gov/russia-spreads-disinformation-to-cover-up-its-use-of-chemical-weapons-in-ukraine} Accessed: 2024-08-15.

\bibitem[{Wasil et~al.(2024{\natexlab{a}})Wasil, Berglund, Reed, Plueckebuam, and Smith}]{wasil2024understanding}
Wasil, A.; Berglund, L.; Reed, T.; Plueckebuam, M.; and Smith, E. 2024{\natexlab{a}}.
\newblock Understanding frontier AI capabilities and risks through semi-structured interviews.
\newblock \url{https://papers.ssrn.com/sol3/papers.cfm?abstract_id=4881729}.

\bibitem[{Wasil et~al.(2024{\natexlab{b}})Wasil, Clymer, Krueger, Dardaman, Campos, and Murphy}]{wasil2024affirmative}
Wasil, A.; Clymer, J.; Krueger, D.; Dardaman, E.; Campos, S.; and Murphy, E. 2024{\natexlab{b}}.
\newblock Affirmative safety: An approach to risk management for high-risk AI.
\newblock \url{https://arxiv.org/pdf/2406.15371}.

\bibitem[{Wasil et~al.(2024{\natexlab{c}})Wasil, Reed, Miller, and Barnett}]{wasil2024verification}
Wasil, A.; Reed, T.; Miller, J.; and Barnett, P. 2024{\natexlab{c}}.
\newblock Verification methods for international AI agreements.
\newblock \url{https://arxiv.org/abs/2408.16074}.

\bibitem[{{Wassenaar Arrangement}(2016)}]{wassenaar_guidelines_2016}
{Wassenaar Arrangement}. 2016.
\newblock Guidelines \& Procedures, including the Initial Elements.
\newblock Public Documents, Vol. 1: Founding Documents.
\newblock Retrieved 6 April 2023.

\bibitem[{{Wassenaar Arrangement}(2024)}]{wassenaar2024about}
{Wassenaar Arrangement}. 2024.
\newblock About Us - FAQ.
\newblock \url{https://www.wassenaar.org/about-us/#faq} Accessed: 2024-08-14.

\bibitem[{{Wassenaar Arrangement Secretariat}(2015)}]{wassenaar2015basic}
{Wassenaar Arrangement Secretariat}. 2015.
\newblock Wassenaar Arrangement on Export Controls for Conventional Arms and Dual-Use Goods and Technologies: Basic Documents.
\newblock \textit{Compiled by the Wassenaar Arrangement Secretariat}.

\bibitem[{Whitehair and Brugger(2001)}]{whitehair2001bwc}
Whitehair, R.; and Brugger, S. 2001.
\newblock BWC protocol talks in Geneva collapse following US rejection.
\newblock \emph{Arms Control Today}, 31(7): 26.

\bibitem[{Woolf(2011)}]{woolf2011new}
Woolf, A.~F. 2011.
\newblock \emph{New START Treaty: Central Limits and Key Provisions}, volume 41219.
\newblock Congressional Research Service.

\end{thebibliography}

\end{document}